\documentclass[11pt]{iopart}

\usepackage[english]{babel}
\usepackage{amssymb}
\usepackage{mathrsfs}
\usepackage[dvips]{graphicx}
\usepackage{epsfig,bm}
\usepackage{txfonts,color}

\begin{document}

\title[Stability of inhomogeneous states]{Stability of inhomogeneous states in 
mean-field models with an external potential}
\author{R.~Bachelard$^1$, F.~Staniscia$^{2,3}$, T. Dauxois$^4$, G. De Ninno$^{1,2}$, S.~Ruffo$^{5}$}

\address{$^1$ University of Nova Gorica, School of applied sciences, Vipavska 11c, SI-5270 Ajdovcina, 
Slovenia
\\ $^2$ Sincrotrone Trieste, S.S. 14 km 163.5, Basovizza (Ts), Italy
\\ $^3$ Dipartimento di Fisica, Universit\`a di Trieste, Italy
\\ $^4$ Laboratoire de Physique de l'\'Ecole Normale Sup\'erieure de Lyon, Universit\'e de Lyon, 
CNRS, 46 All\'ee d'Italie, 69364 Lyon c\'edex 07, France
\\ $^5$ Dipartimento di Energetica ``Sergio Stecco", Universit\`a di Firenze and INFN, via S. Marta 3, 
50139 Firenze, Italy
}
\ead{bachelard.romain@gmail.com}

\date{}

%\tableofcontents

\begin{abstract}
The Vlasov equation is well known to provide a good description of the dynamics of mean-field
systems in the $N \to \infty$ limit. This equation has an infinity of stationary states and
the case of {\it homogeneous} states, for which the single-particle
distribution function is independent of the spatial variable, is well characterized analytically. On the other hand, the inhomogeneous case often requires some approximations for an analytical treatment: the dynamics is then best treated in action-angle variables, and the potential generating inhomogeneity is generally very complex in these new variables. We here treat analytically the linear stability of toy-models where the inhomogeneity is created by an external field. Transforming the Vlasov equation into
action-angle variables, we derive a dispersion relation that we accomplish to solve for both
the growth rate of the instability and the stability threshold for two specific models: the
Hamiltonian Mean-Field model with additional asymmetry and the mean-field $\phi^4$ model. The results are compared with 
numerical simulations of the $N$-body dynamics. When the {\it inhomogeneous} state is stationary stable, 
we expect to observe in the $N$-body dynamics Quasi-Stationary-States (QSS), whose lifetime diverges 
algebraically with $N$.
\end{abstract}

\maketitle
%\tableofcontent/s

\section{Introduction}

Long-range forces can be found in a wide variety of physical systems, including self-gravitating systems, 
Coulomb systems, wave-plasma interactions and two-dimensional hydrodynamics. The interest in studying long-range
forces has been revived in the last decade, not only because of the broad domain of physical systems involving 
such forces, but also because of the presence of unusual phenomena, both at equilibrium and out of equilibrium. 
Let us mention negative specific heat, temperature jumps, broken ergodicity and quasi-stationary states.
Reviews and books have been recently published in this field
\cite{Campa,Leshouches1,Assisi,Leshouches2,bouchet,chavareview}.

A particular, but interesting,
case is the one of {\it mean-field} interactions, for which each particle is
directly coupled to all the others with equal strength, whatever their distance. Although this is an idealization, it serves as a useful approximation and 
appears, in addition, to give at least the good trend. 
Moreover, there are physical situations in which particles are all in interaction
via a field, whose dynamics is in turn determined by the motion of the particles
themselves: this is for example the case of wave-particle interactions in plasmas~\cite{eebook},
Free Electron Lasers~\cite{colsonbonifacio}, Collective Atomic Recoil Lasers~\cite{CARL} and
Traveling Wave Tubes~\cite{TWT}. This self-consistent effect can also be obtained
in systems composed only of particles by introducing a coupling to an {\it order
parameter}, as it is done for the Hamiltonian Mean Field (HMF) model \cite{inagaki,pichon,hmf95},
which has been widely studied in recent years as a paradigm for systems
with long-range interactions \cite{Campa} .

The kinetics of models with $N$ particles and only mean-field interactions is exactly
described, in the infinite $N$ limit, by the Vlasov equation \cite{MesserSpohn,Braun}. This
equation exhibits an infinity of stationary solutions and its dynamical evolution
starting from a generic initial state can be extremely complex. Focusing on stationary
states, their stability has been studied using different methods, but mainly by restricting
the analysis to {\it homogeneous} stationary states, that are characterized by a single-particle 
distribution function which is independent
of the spatial variable. These states are of major interest in kinetic theory,
because they often constitute the ``supposed'' physical equilibrium state.  For instance a globally neutral
plasma has an equilibrium which is also locally neutral, giving a {\it homogeneous}
charge distribution. If perturbed, this state is expected to be stable, showing a relaxation
back to the homogeneous state ruled by Landau damping \cite{Landaukinetic,Nicholson,Balescubook}.
This phenomenology is also observed in the HMF model \cite{yamaguchi04}, for which the homogeneous
state is stable above a given energy threshold, which depends on the initial momentum
distribution.

However, below this energy, the homogeneous state is unstable and one
observes a dynamical evolution towards inhomogeneous states, whose stability properties are
much more difficult to determine. Inhomogeneous states 
appear for example in gravitational dynamics \cite{BinneyTremaine}, because of the attractive nature of the Newton
force. Their stability has been studied in the context of the Vlasov equation, yet the necessity to resort to action-angle variables~\cite{brink} makes the problem analytically tricky. Apart from numerical approaches (see e.g.~\cite{bertin}), one can project the dynamics onto a Fourier basis, yet at a cost of performing infinite sums~\cite{kaljnas,weinberg}; then, only a truncation can yield tractable results. Such technique was also used in the context of plasmas~\cite{schindler,krall}, where the waves often generate inhomogeneous states; expanding the dynamics along modes, such as Hermite polynomials~\cite{camporeale}, requires anyhow a truncation in the sums. Analytical results were also obtained on BGK modes, whose stability properties were connected, in the small inhomogeneity limit, to those of homogeneous states~\cite{guo,manfredi}. Later on, the unstable nature of periodic BGK modes under specific perturbations was rigorously shown~\cite{lin2001,lin2005}, but the problem remains open for other types of systems and perturbations. More recently, some general criteria were proposed to derive the stability of inhomogeneous states~\cite{campachacha,chachacha}.

Some toy models were also studied whose states are naturally inhomogeneous: this is typically the case of systems when an external potential is present in addition to the self-consistent one~\cite{phi4,Hahn,Touchette,HMFC2}.
A first interesting model~\cite{phi4,Hahn,Touchette,Zwanzig} is the mean-field $\varphi^4$ model: an Ising-like spin variable
is represented by a scalar field in one dimension, acted upon externally by a double-well potential
which selects two states; the mean-field term of the Hamiltonian is a quadratic coupling
of the scalar field at two different lattice sites.
A second interesting model is a generalized version of the Hamiltonian Mean-Field (HMF) model to which an anisotropic external potential is 
added~\cite{HMFC2} .

In this paper we focus the above mentioned toy models, and show that one can treat {\it exactly} the stability of inhomogeneous
states.
The Vlasov equation will be rewritten in action-angle variables \cite{chacha07,chacha10,barre10}
and we will focus on those inhomogeneous stationary states whose single-particle distribution function 
does not depend on the angle variable, i.e. those that are homogeneous in angle. We will derive a general stability 
criterion which, besides giving the value of the threshold energy (action) at which these stationary states 
destabilize, will allow us to obtain the growth-rate of the instability.

In Section~\ref{sec:theory} we will introduce and discuss the Vlasov equation in action-angle variables
and we will derive the stability condition for inhomogeneous states and for generic mean-field and external potentials.
In Sections~\ref{sec:HMF-Cos2} and \ref{sec:phi4} we shall apply the general method introduced in 
Section~\ref{sec:theory} to the specific cases of the anisotropic HMF model and of the mean-field $\varphi^4$ model,
deriving explicit analytical expressions for the stability threshold and for the growth rate of the instability. These
theoretical predictions will be then compared with numerical simulations performed with $N$-body
Hamiltonians. Finally, in Section~\ref{sec:ccl}, we will draw some conclusions and we will discuss some perspectives 
of this work.

\section{The Vlasov equation in action-angle variables and the stability relations 
\label{sec:theory}}

Let us consider $N$ particles in one-dimension whose positions and momenta are $(q_j,p_j)$, $j=1,\dots,N$.
They interact through the two-body (symmetric) potential $v(q_j,q_k)$ and, in addition, each particle 
is trapped into the external potential $W(q_j)$. Hamilton's equations for such a system are
\begin{eqnarray}
\dot{q}_j &=& p_j, \label{eq:pjdot}
\\ \dot{p}_j &=& -W'(q_j)-  \partial_{q_j} V\left[\{q_k\}\right](q_j), \label{eq:qjdot}
\end{eqnarray}
where $V\left[\{q_k\}\right](q_j)=(1/N)\sum_k v(q_j,q_k)$ stands for the mean-field potential
acting on particle $j$. 
The $1/N$ term is a rescaling factor~\cite{kac} which allows one to perform the mean-field limit discussed
in Refs.~\cite{MesserSpohn,Braun}. The prime will denote, from now on, the derivative with respect to the position 
variable~$q$. Eqs.~(\ref{eq:pjdot}) and~(\ref{eq:qjdot}) can be derived from the following Hamiltonian
\begin{equation}
H=\sum_j \left(\frac{p_j^2}{2} + W(q_j) +\frac{1}{2} V\left[\{q_k\}\right](q_j) \right)~,\label{eq:HNbodymf}
\end{equation}
where the $(q_j,p_j)$ are couples of canonically conjugated variables. Let us introduce the so-called {\it empirical measure}
\begin{equation}
f(q,p,t)=\frac{1}{N} \sum_{j=1}^N \delta (q-q_j(t)) \delta (p-p_j(t))~.
\end{equation}
It can be shown \cite{Braun} that, in the $N \to \infty$ limit, the single-particle distribution function 
$f(q,p,t)$ obeys 
the following Vlasov equation 
\begin{equation}
\partial_t f+p \partial_q f-\left(W'(q)+ V'[f](q) \right)\partial_p f=0~, \label{eq:Vlmf}
\end{equation}
where
\begin{equation}
V[f](q,t)=\iint \mbox{d}q'\mbox{d}p'\ f(q',p',t)v(q,q'), \label{defpotential}
\end{equation}
is the averaged mean-field potential.
One can also show that the $N$-body dynamics is well described by the Vlasov equation over times 
that are at least of order $\ln N$~\cite{Braun}. This makes the Vlasov framework a natural one to study such 
systems when a large number of particles is involved. 

The Vlasov equation can also be written in Hamiltonian form using the following functional
\begin{equation}
\label{functionalH}
H[f]=\iint \mbox{d}q\mbox{d}p\ f(q,p,t) \left(\frac{p^2}{2} + W(q) +\frac{1}{2} V[f](q) \right)~.
\end{equation}
After having introduced the appropriate Poisson brackets for the functionals $A[f]$ and $B[f]$
\begin{equation}
\{A,B\}=\iint \mbox{d}q\mbox{d}p\ f(q,p,t) \left( \frac{\partial }{\partial p}\frac{\delta A}
{\delta f}\frac{\partial }{\partial q}\frac{\delta B}{\delta f}-\frac{\partial }{\partial q}\frac{\delta A}
{\delta f}\frac{\partial }{\partial p}\frac{\delta B}{\delta f}\right)~,\label{eq:bracket}
\end{equation}
the dynamics of $A[f]$ is given by
\begin{equation}
\partial_t A =\{H,A\}. \label{eq:dynobs}
\end{equation}
If one rewrites the single particle distribution function in the functional form  
$f(q,p,t)=\iint \mbox{d}q'\mbox{d}p'\ f(q',p',t)\delta(q-q')\delta(p-p')$, one obtains the evolution equation
\begin{eqnarray}
&&\partial_t f(q,p,t) +\frac{\partial h}{\partial p}\frac{\partial f(q,p,t)}{\partial q}
-\frac{\partial h}{\partial q}\frac{\partial f(q,p,t)}{\partial p} \nonumber\\
&&=\partial_t f(q,p,t)+\{h[f](q,p),f(q,p,t)\}=0 \label{eq:VlasovHam}
\end{eqnarray}
where $h[f](q,p)=p^2/2+W(q)+V[f](q)$ and the brackets are now the standard Poisson brackets.
This equation is nothing but the Vlasov equation~(\ref{eq:Vlmf}). 

It is straightforward to check that the Boltzmann-Gibbs equilibrium distribution 
$f_{BG}(q,p)=Z^{-1}\exp (-\beta h(q,p))$, with $\beta$ an arbitrary constant and $Z$ a normalization constant, 
is a stationary solution of this equation (i.e. $\partial_t f_{BG}=0$). In fact, all distributions
that depend on $(q,p)$ only through $h$ are stationary. The existence of an infinity of stationary 
distributions is actually responsible for the peculiar out-of-equilibrium regimes in which $N$-body 
long-range systems get trapped over very long times \cite{Campa}. More specifically, starting from a 
generic unstable distribution, a long-range system typically relaxes towards a ``quasi-stationary'' 
state, which can be significantly different from Boltzmann-Gibbs equilibrium. Quasi-Stationary
States (QSS) can be interpreted as stable stationary states of the Vlasov equation in the $N \to \infty$ limit.
The relaxation to statistical equilibrium occurs on much longer time scales, that were observed to diverge either algebraically
\cite{yamaguchi04} or logarithmically \cite{HMFC2} with $N$, depending on whether the ``quasi-stationary'' 
state corresponds to a stable or an unstable stationary state of the Vlasov equation. Relaxation to
equilibrium is not due to collisions but due to finite-$N$ effects (also called ``granularity''), which can be 
modeled by convenient kinetic equations, like Landau or Lenard-Balescu equations~\cite{Campa,Nicholson,Balescubook}. 
Stable stationary states of the Vlasov equation are therefore of paramount importance in order
to understand the dynamics of long-range systems. It is therefore crucial to determine the general conditions for 
stationarity and stability, for both homogeneous and inhomogeneous states.

Let us consider the stationary state $f_0(q,p)$. If one focuses on the Lagrangian trajectory of a single particle, 
one immediately realizes that it is a constant energy trajectory of the energy functional 
\begin{equation}
h[f_0](q,p)=\frac{p^2}{2}+W(q)+V[f_0](q),
\end{equation} 
which is a straightforward consequence of Eqs.~(\ref{eq:pjdot}) and~(\ref{eq:qjdot}). 
Hence, it is convenient to cast the dynamics into the appropriate variables associated with this trajectory, namely the 
``action-angle" variables
\begin{eqnarray}
J(h)&=&\frac{1}{2\pi}\oint p(h,q')dq'=\frac{1}{2\pi}\oint \sqrt{2\left(h-W(q')-V[f_0](q') \right)} dq'\label{eq:j}
\\ \phi &=&\omega\int_{0}^q \frac{dq'}{\sqrt{2\left(h-W(q')-V[f_0](q') \right)}},\label{eq:phi}
\end{eqnarray}
where the frequency $\omega$ is given by
\begin{equation}
\omega=\left(\frac{1}{2\pi}\oint \frac{dq'}{\sqrt{2\left(h-W(q')-V[f_0](q') \right)}}\right)^{-1}
=\frac{\partial h}{\partial J}.
\end{equation}
It is important to note that the conjugate variables $(J,\phi)$ are not action-angle {\it stricto sensu}: Since Vlasov dynamics is infinite dimensional and only a specific set of conserved quantities can be typically 
identified (e.g. the Hamiltonian, total momentum, the Casimirs $\iint \mbox{d}q\mbox{d}p\ C(f(q,p))$, with
$C$ an analytic function), its integrability is not generic~\cite{gibbons}. The term action-angle variables comes 
from the fact that the dynamics of a Lagrangian test-particle is integrable if the single-particle
distribution function is stationary. Indeed, for a stationary distribution $f_0$, the potential $V[f_0]$ is 
constant in time. Therefore, the dynamics of the test-particle is that of a one--degree--of--freedom system with the 
associated conserved quantity $h[f_0]$, hence integrable. A dependence of the potential on time caused by a 
non-stationary distribution $f(q,p,t)$ would introduce an extra $1/2$ degree of freedom, 
thus breaking integrability {\it a priori}.

In this single particle framework and for stationary distributions, the energy $h$ depends only on the action $J$, 
so that a particle evolves on a trajectory of constant ``action'' $J$ at the constant action-dependent angular speed 
$\dot{\phi}=\partial_J h(J)=\omega(J)$. 
The change of variables $(q,p)\rightarrow (\phi,J)$ being canonical, the corresponding Poisson brackets,
which apply to functions of the phase-space, are equivalent
\begin{equation}
\{a,b\}_{q,p}=\partial_p a \partial_q b-\partial_q a\partial_p b=\{a,b\}_{\phi,J}=\partial_J a 
\partial_\phi b-\partial_\phi a\partial_J b.\label{eq:redbra}
\end{equation}

Using this equivalence and the condition $\partial_\phi h=0$, the Vlasov equation (\ref{eq:Vlmf}) for $f_0$ can be 
recast in the following form
\begin{equation}
\partial_J h(J) \partial_\phi f_0=\omega(J) \partial_\phi f_0=0~.\label{eq:statio}
\end{equation}
Hence, the stationarity condition, $\partial_t f_0=0$, leads to $f_0=f_0(J)$. This means 
in particular that the stationary distributions are those that are {\it homogeneous in angle}, with {\it any} distribution 
in action~$J$. Such a result highlights the relevance of action-angle variables for the analysis of Vlasov stationary dynamics, 
but also for the study of QSS.

We shall now consider a perturbation $\delta f$ around $f_0$, that is $f(\phi,J)=f_0(J)+\delta f(\phi,J)$. 
The linearity of the potential $V$ with respect to the distribution, as emphasized by its definition in Eq.~(\ref{defpotential}), 
implies that $V[f]=V[f_0]+V[\delta f]$. 
Using property (\ref{eq:redbra}) for the Vlasov equation~(\ref{eq:Vlmf},\ref{eq:VlasovHam}) and neglecting second-order 
terms in $\delta f$ leads to the {\it linearized Vlasov equation}
\begin{equation}
\label{eq:Vlasovlin}
\partial_t \delta f +\omega(J)\partial_\phi \delta f-(\partial_p f_0)V'[\delta f](\phi,J)=0~,
\end{equation}
where the factor $\partial_p f_0$ should be expressed in terms of $(\phi,J)$ and the derivative of $V$ is
with respect to $q$ and then it is also expressed in terms of $(\phi,J)$.
The study of this equation in full generality would imply the solution of an initial value problem using a Laplace-Fourier 
transform and then a transformation back to action-angle variables using a Bromwich contour \cite{Nicholson,Balescubook}.
We will be here less ambitious and we will focus on the study of an eigenmode 
$\delta f(\phi,J;t)=e^{\lambda t}\bar{f}(\phi,J)$ with the eigenvalue $\lambda$ determining the stability properties. 
Inserting this ansatz solution in Eq.~(\ref{eq:Vlasovlin}) one gets
\begin{equation}
(\lambda +\omega(J)\partial_\phi) \overline{f}-(\partial_p f_0)V'\left[\overline{f}\right](\phi,J)=0.
\end{equation}
Assuming a non-zero $\omega$ (the frequency $\omega$ typically only vanishes on the separatrices of the single particle phase-space), 
the above equation turns into
\begin{equation}
\partial_\phi \left(e^{\lambda \phi/\omega(J)}\bar{f}\right)-\frac{e^{\lambda{\phi}/{\omega(J)}}}{\omega(J)}(\partial_p f_0)V'\left[\overline{f}\right](\phi,J)=0.
\end{equation}
After integration over the angle $\phi$, and assuming that the integration constant vanishes, one gets
\begin{equation}
\bar{f}-\frac{e^{-\lambda \phi/ \omega(J)}}{\omega(J)}\int_{0}^{\phi}\mbox{d}
\phi' e^{\lambda{\phi'}/{\omega(J)}}(\partial_p f_0)(\phi',J)V'\left[\overline{f}\right](\phi',J)
=0.\label{eq:dispt1}
\end{equation}
This equation can be fully cast into action-angle variables using the following relation
\begin{equation} 
\frac{\partial f_0}{\partial p}(q,p)=\frac{\partial J}{\partial p} \frac{\partial f_0(J)}{\partial J}
=\frac{\partial h}{\partial p} \frac{\partial J}{\partial h} f_0'(J)=\frac{p}{\omega}f_0'(J)~,
\end{equation}
which, inserted into Eq.~(\ref{eq:dispt1}), results in the following {\it dispersion relation}
\begin{equation}
\bar{f}-f'_0(J)\frac{e^{-\lambda\phi/\omega(J)}}{\omega^2(J)}\int_{0}^{\phi}\mbox{d}\phi'  
p(\phi',J)\,e^{\lambda{\phi'}/{\omega(J)}} \, V'\left[\overline{f}\right](\phi',J)=0.\label{eq:fbar}
\end{equation}
It is convenient to express the integral in this latter equation in terms of the position variable
$q'$. Indeed, using Eq.~(\ref{eq:phi}), the differential $\mbox{d}\phi'$ can be calculated as a function of $q'$ at constant 
action $J$, which means along a single-particle trajectory. One gets
\begin{equation}
\mbox{d}\phi'=\frac{\omega \mbox{d}q'}{\sqrt{2\left(h-W(q')-V[f_0](q') \right)}}=\frac{\omega}{p} \mbox{d}q'\label{eq:dphi}~,
\end{equation}
which allows one to put Eq.~(\ref{eq:fbar}) into the following form
\begin{equation}
\bar{f}-f'_0(J)\frac{e^{-\lambda{\phi}/{\omega(J)}}}{\omega(J)}\int_{0}^{q}\,e^{\lambda{\phi'}/{\omega(J)}}\, 
V'\left[\overline{f}\right](q')\mbox{d}q'=0,\label{eq:fbar2}
\end{equation}
in which the integral is performed at constant action~$J$. The interest of this alternative formula is that 
it may be easier to solve in some cases. In particular, if one focuses on the stability threshold, given by 
taking $\lambda=0$, the integral over $q'$ can be solved straightforwardly and Eq.~(\ref{eq:fbar2}) can be rewritten
as
\begin{equation}
\bar{f}=\frac{f'_0(J)}{\omega(J)} V\left[\overline{f}\right](q),\label{eq:stabthreshold}
\end{equation}
where $q$ is, in general, a function of both action and angle.

Since all functions in angle are $2\pi$-periodic, it is common to project the dispersion relation in a Fourier
base ~\cite{chacha07,barre10}. However, since in Eq.~(\ref{eq:fbar}) both the $p$ term and the potential 
$V[\bar{f}]$ have generically a non trivial dependence on the angles, one ends up with expressions where all
Fourier modes are coupled. The modes are decoupled only when momentum does not depend on angle, which
is the case of homogeneous states, for which momentum coincide with action (modulo a sign).

In what follows we will discuss a method which allows us to compute the stability threshold and the growth
rate $\lambda$ without resorting to a Fourier expansion. The method is, however, not generic and its application
depends on the specific form of the interaction potential. We will therefore discuss separately two examples.

\section{The HMF model with additional asymmetry \label{sec:HMF-Cos2}}

Introduced in Ref.~\cite{HMFC2}, the HMF model with additional cosine on-site potential is a generalization of the 
paradigmatic HMF model~\cite{inagaki,pichon,hmf95}. Besides the mean-field term $v(q_j,q_k)=-\cos{(q_j-q_k)}$, an external
potential $W$ of amplitude $\kappa$ is present  
\begin{equation}
W(q_j)=\kappa \cos^2{q_j}~.
\end{equation} 
The Hamiltonian (\ref{functionalH}) reads
\begin{equation}
\label{eq:HMFcos2}
H[f]=\iint \mbox{d}q\mbox{d}p\,f(q,p) \left[\frac{p^2}{2} + \kappa \cos^2q - \frac{1}{2} 
\iint \mbox{d}q'\mbox{d}p' \,f(q',p')\cos{(q-q')}\right]~.
\end{equation}
At variance with the HMF model, the spatially-homogeneous state is no longer a stationary state of the 
Vlasov equation, due to the presence of the on-site potential.

Using formula~(\ref{eq:fbar}), one easily gets the dispersion relation for this model
\begin{equation}
\bar{f}-f'_0(J)\frac{e^{\lambda{\phi}/{\omega(J)}}}{\omega^2(J)}\int_{0}^{\phi}\mbox{d}\phi'\,p(\phi',J)\, e^{\lambda{\phi'}/{\omega(J)}} \left[M_x[\bar{f}]\sin(q(\phi',J))-M_y[\bar{f}]\cos(q(\phi',J))\right]=0,\label{eq:disphmfc2}
\end{equation}
where 
\begin{equation}{\bf M}[f]=M_x[f]+i M_y[f]=\iint \mbox{d}q \mbox{d}pf(q,p) \cos q+i\iint \mbox{d}q \mbox{d}p 
f(q,p) \sin q
\end{equation} 
stands for the magnetization. For the sake of simplicity,  $q$, $q'$ and $p'$ will respectively refer 
to $q(\phi,J)$, $q(\phi',J)$ and $p(\phi',J)$ in the remaining of this section. 
Equation (\ref{eq:disphmfc2}) can be solved by multiplying each term by either $\cos q$ or $\sin q$, 
and then integrating over phase-space. 
One gets the following equations
\begin{eqnarray} 
M_x[\bar{f}]\left(1-I_{X,Y}^\lambda[f_0]\right)&+M_y[\bar{f}]I_{X,X}^\lambda [f_0]&=0,\label{eq:systc2b}
\\ -M_x[\bar{f}]I_{Y,Y}^\lambda[f_0]&+M_y[\bar{f}]\left(1+I_{Y,X}^\lambda[f_0]\right)&=0,\label{eq:systc2}
\end{eqnarray}
where
\begin{equation}
I_{X,Y}^\lambda[f_0]=\int dJ \frac{f'_0(J)}{\omega(J)}\oint \mbox{d}\phi\, e^{-\lambda{\phi}/{\omega(J)}}
\,X(q)\int_{0}^q \mbox{d}q'\,e^{\lambda{\phi'}/{\omega(J)}}\,Y(q'),\label{eq:IXY}
\end{equation}
and the label $X$ (resp. $Y$) stands for the $\cos$ (resp. $\sin$) function. The integral $\oint$ is performed 
over a single-particle trajectory. 

Inhomogeneous stationary states of the Vlasov equation correspond to solutions of the 
linear system of equations (\ref{eq:systc2b}-\ref{eq:systc2}) with non vanishing $(M_x,M_y)$.
They can be found only when the determinant vanishes. This condition allows to rewrite the
dispersion relation in the form
\begin{equation}
\left(1-I_{X,Y}^\lambda[f_0]\right)\left(1+I_{Y,X}^\lambda[f_0]\right)+I_{Y,Y}^\lambda[f_0]
I_{X,X}^\lambda [f_0]=0.\label{eq:disprelhmfc2}
\end{equation}
The numerical resolution of this equation can be performed by using the explicit expressions of the 
action-angle coordinates~\cite{lichtenberg}, for a particle of energy $h$ and position $q$
\begin{eqnarray}
J_{in}(h)=\frac{2\sqrt{2\kappa}}{\pi}\ \left[\mathscr{E}\left(\frac{h}{\kappa}\right)-\left(1-\frac{h}{\kappa}\right)\mathscr{K}\left(\frac{h}{\kappa}\right)\right]\qquad& \phi_{in}(q,h)=\frac{\pi}{2}\sqrt{\frac{\kappa}{h}} \frac{\mathscr{F}\left(q,{h}/{\kappa}\right)}{\mathscr{K}\left({h}/{\kappa}\right)}
\\ J_{out}(h)=\frac{2\sqrt{2h}}{\pi}\ \mathscr{E}\left(\frac{\kappa}{h}\right)& \phi_{out}(q,h)=\frac{\pi}{2}\frac{\mathscr{F}\left(q,\kappa/h\right)}{\mathscr{K}\left(\kappa/h\right)},
\end{eqnarray}
where the label $in/out$ stands for inside/outside of the separatrix of the potential $\kappa\cos^{2} q$, 
while $\mathscr{E}$, $\mathscr{K}$ and $\mathscr{F}$ are elliptic integrals of the first kind.

\begin{figure}[!ht]
\centerline{
$\begin{array}{ccc}
\epsfig{figure=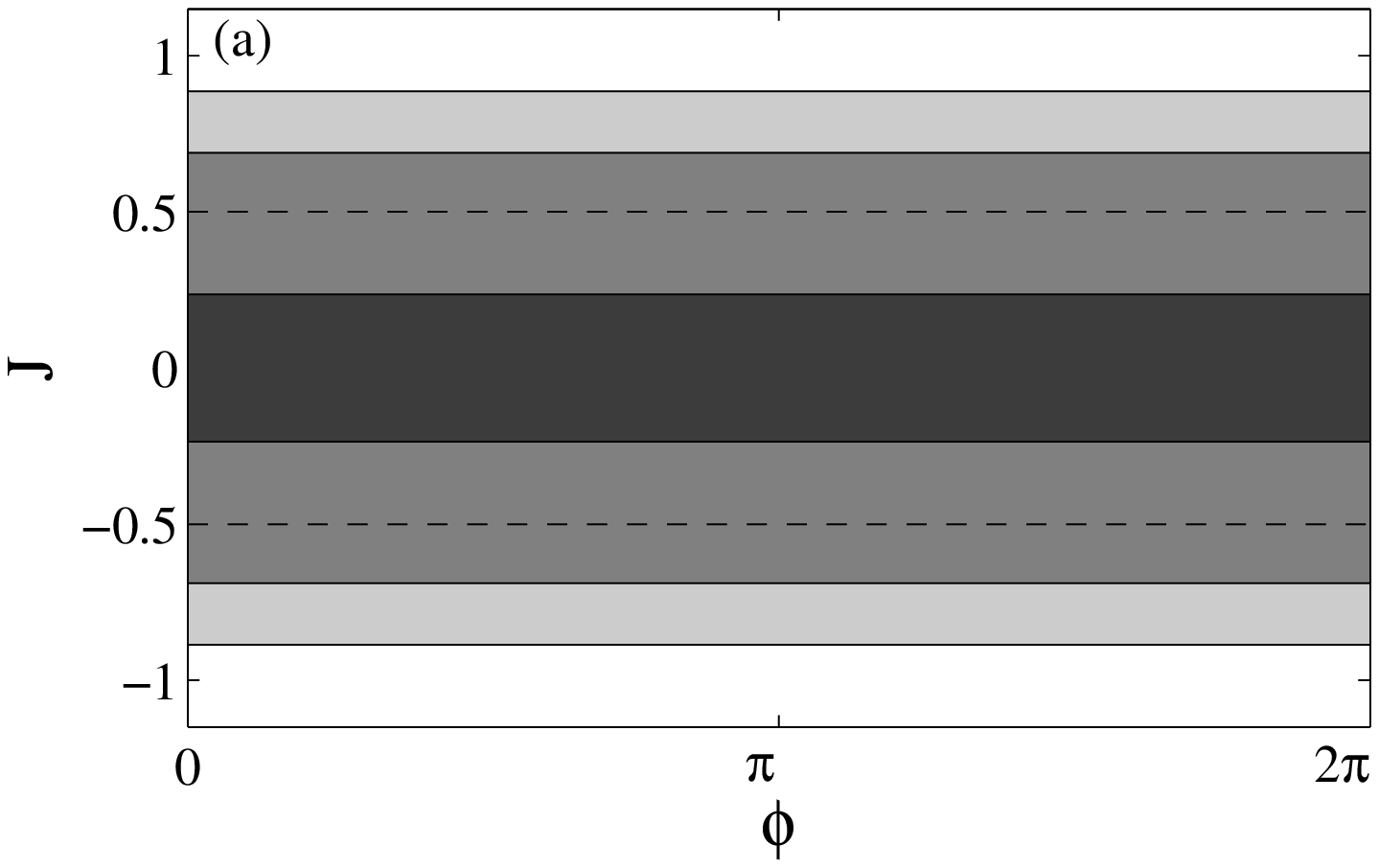,width=8cm} & \epsfig{figure=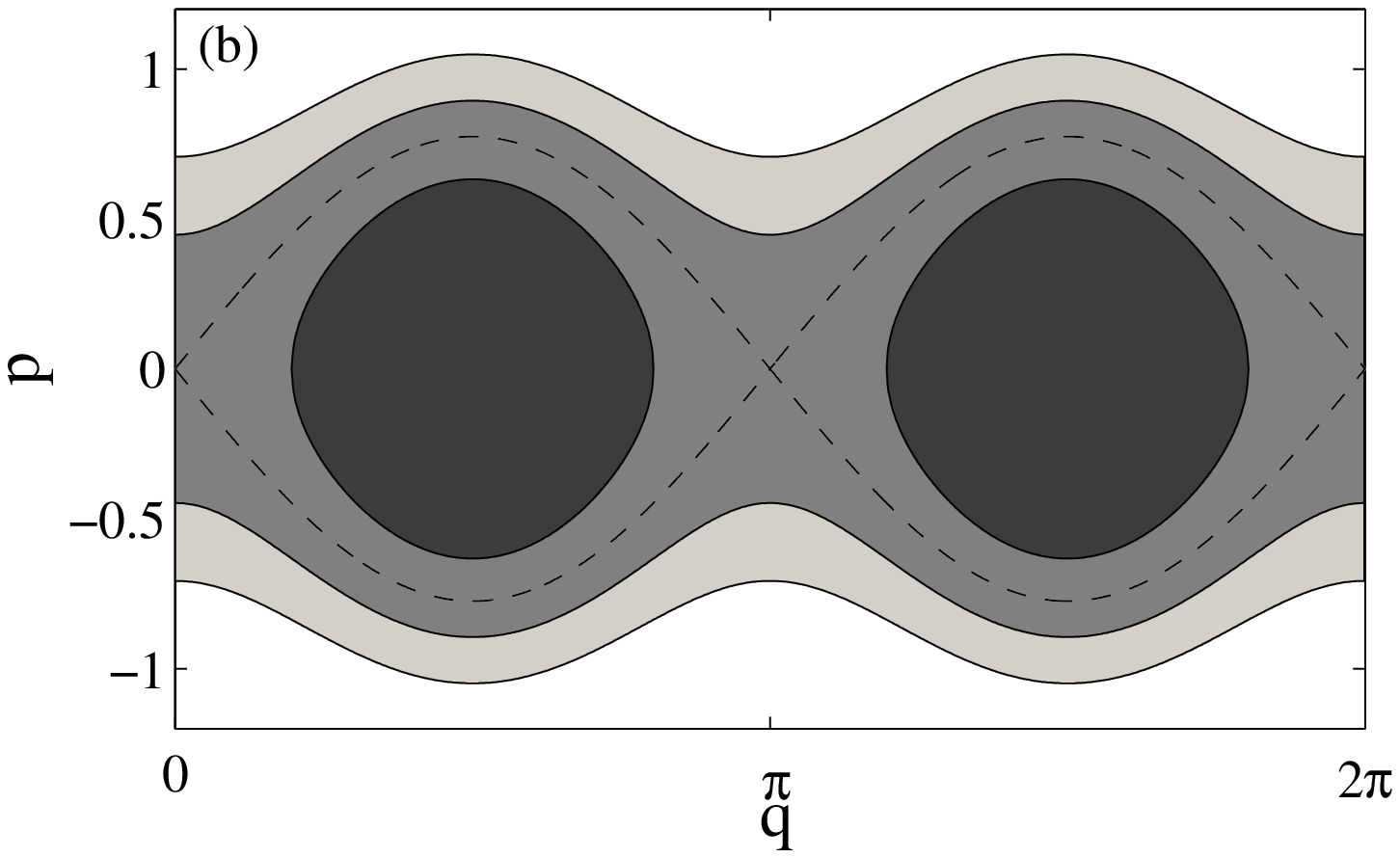,width=8cm}  
\end{array}$}
\caption{Waterbags in action-angle (panel a) and in $(q,p)$ space (panel b). The waterbags
have increasing boundary energies $U=0.2$, $0.4$ and $0.55$ and they are represented by filled contours
of lighter and lighter grey as the energy is increased. The dashed line corresponds to the separatrix, which has
energy $U_{s}=0.3$ and action $J_s=0.5$. \label{fig:waterbag}
}
\end{figure}

\begin{figure}[!ht]
\centerline{\epsfig{figure=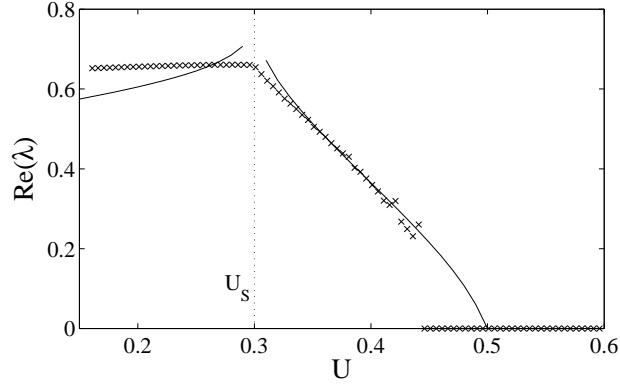,width=9cm}}
\caption{Growth rate Re$(\lambda)$ (full line) of the instability of the inhomogeneous waterbag states obtained by 
solving Eq.~(\ref{eq:disprelhmfc2}) for waterbags with boundary energy $U$. The crosses are the results of 
numerical simulations of the $N$-body Hamiltonian. The agreement between theory (which decribes the $N \to \infty$ 
limit) and numerics (which is performed at $N=3 \times 10^5$) is reasonably good apart from the region near the 
separatrix energy $U_s=0.3$ and the one near the critical energy $U_c=0.498$, which is theoretically determined by
solving Eq.~(\ref{eq:stabthres}). \label{fig:lyaphmfc2}
}
\end{figure}

In order to compute the growth rate Re$(\lambda)$ from Eq.~(\ref{eq:disprelhmfc2}) it is necessary to choose
a specific unperturbed stationary distribution $f_0(J)$. We here consider ``waterbag'' distributions in
action-angle space that are homogeneous in angle: these are two-level distributions, which are nonzero and homogeneous between two lines of constant action $J=J_1$ and $J=J_2$
\begin{equation}
f_0(J)=\frac{1}{2\pi(J_2-J_1)}\left(\Theta(J-J_1)-\Theta(J-J_2)\right),\label{eq:wb}
\end{equation}
where the first factor guarantees the normalization of the density $f_0$, while $\Theta$ is the Heaviside 
step function.
Morever, we here focus on waterbags delimited by a given energy $U$, i.e. we consider all trajectories 
with energies $h\leq U$ (so that $J_2=J_2(U)$ and $J_1=0$), such as those represented in 
Fig.~\ref{fig:waterbag}(a) and (b).
It is interesting to remark that, since the change of variables $(q,p)\leftrightarrow (\phi,J)$ is canonical, 
$f_0(q,p)$ is also a two-step distribution with the boundary given by the curve $h(q,p)=U$.
It should be pointed out that, altough the action fixes the energy univocally, a trajectory of
given energy is always splitted in two: those with positive and negative momentum $p$ for $U>U_s=0.3$,
the separatrix energy, and the ones with $0 < q < \pi$ and $\pi < q < 2\pi$ for $U<U_s$. This has 
the consequence that, when performing integrations over the action-angle space, the two trajectories
give separate contributions. Related to this remark is for example the evaluation of the normalization of $f_0$: 
the total area of the waterbag is indeed $2 \times 2 \pi (J(U)-J(0))=4 \pi J(U)$.

For the waterbag initial conditions, the integral in Eq.~(\ref{eq:IXY}) reads
\begin{eqnarray}
I_{X,Y}^\lambda[f_0]=&\frac{1}{2\pi(J_2-J_1)} \Bigg[\frac{1}{\omega(J_1)}\oint \mbox{d}\phi\, 
e^{-\lambda{\phi}/{\omega(J_1)}}X(q)\int_{0}^q \mbox{d}q'\,e^{\lambda{\phi'}/{\omega(J_1)}}Y(q')\nonumber
\\ &\hskip 1.8truecm -\frac{1}{\omega(J_2)}\oint \mbox{d}\phi \,e^{-\lambda{\phi}/{\omega(J_2)}}X(q)\int_{0}^q 
\mbox{d}q'\,e^{\lambda{\phi'}/{\omega(J_2)}}Y(q')\Bigg]~.
\label{eq:newIXY}
\end{eqnarray}

The numerical solution of Eq.~(\ref{eq:disprelhmfc2}), using Eq.~(\ref{eq:newIXY}), are then compared 
with the result of simulations performed with the $N$-body Hamiltonian using a sixth-order integration 
scheme~\cite{atela} with time step $0.1$.
Figure~\ref{fig:lyaphmfc2} shows the growth rate Re$(\lambda)$ obtained theoretically
(full line) as a function of the boundary energy $U$. The growth rate is determined numerically by fitting 
an exponential to the short-time increase of the magnetization. One notices the existence of a threshold
energy $U_c=0.498$ (determined more precisely in the following), which separates a region where the waterbag
is stable ($U>U_c$) from one where the waterbag is unstable ($U<U_c$, Re$(\lambda)>0$). When the waterbag
is stable, the $N$-body dynamics shows a QSS regime with zero magnetization but with an {\it inhomogeneous}
distribution of particles in the $q$ spatial coordinate.
Let us remark that the theoretical results shows a divergence of Re$(\lambda)$ at the separatrix energy
$U=U_s=0.3$ where the frequency $\omega(J_s)=0$: this divergence is not reproduced by the $N$-body dynamics.
Moreover, in the $N$-body dynamics, the threshold energy is found to be around $U \approx 0.44$, well below
the theoretical value. Indeed, in the energy region $0.44 < U < U_c$ the growth of the magnetization is spoiled by 
finite-$N$ effects, and its exponential character is not clear any more. However, the energy $U_c$ is really
the one where we numerically observe a destabilization of the zero magnetization state.

The critical energy $U_c$ beyond which the waterbags become stable can be explicitly derived using 
Eq.~(\ref{eq:disprelhmfc2}) and by imposing  $\lambda=0$. Let us first note that, in this equation, the last term 
vanishes, since both $I_{X,X}^0 [f_0]$ and $I_{Y,Y}^0 [f_0]$ yield an integral of $\sin q\cos q$ 
over a trajectory. Consequently, the product $(1-I_{X,Y}^0[f_0])(1+I_{Y,X}^0[f_0])$ should be zero. 
Then, considering that $|p|=\sqrt{2U_c}\sqrt{1-(\kappa/U_c)\cos^2 q}$, integrating over $q'$, and using 
Eq.~(\ref{eq:dphi}) and then Eq.~(\ref{eq:j}), we finally get
\begin{eqnarray}
I_{X,Y}^0[f_0]&=&\frac{2}{4\pi \omega_c J_c} \oint \mbox{d}\phi \cos^2q \\
&=&\frac{1}{2\pi J_c} \oint \mbox{d}q \frac{\cos^2q}{p}  \\
&=&\frac{1}{2}\frac{\oint  |\mbox{d}q|\frac{\cos^2q}{\sqrt{U_c-\kappa \cos^2q }}}{\oint  |\mbox{d}q|\sqrt{U_c-\kappa \cos^2 q} },  \label{eq:Ics}
\\ I_{Y,X}^0[f_0]&=&-\frac{1}{2}\frac{\oint |\mbox{d}q| \frac{\sin^2 q}{\sqrt{U_c-\kappa \cos^2q}} }{\oint  |\mbox{d}q|\sqrt{U_c-\kappa \cos^2 q} }. \label{eq:Isc}
\end{eqnarray}
Let us explain the meaning of the uncommon notation $|\mbox{d}q|$. When integrating over segments of the
single-particle trajectory where $p$ is negative, $q$ decreases. Thus, both $\mbox{d}q$ and $p$ are negative, 
so that their ratio or product is positive. The use of the differential $|\mbox{d}q|$ allows us   
to unify notation for both the cases in which  $p$ and $\mbox{d}q$ are positive or negative.
The coefficient $2$ in front of the first integral originates from the double boundary of the waterbag, 
be it inside or outside the separatrix. It can be shown that both expressions~(\ref{eq:Ics}) and (\ref{eq:Isc}) 
are strictly decreasing functions of $U_c$. Moreover, integral~(\ref{eq:Ics}) tends to one in the 
$U_c\rightarrow \kappa$ limit, so that $1-I_{X,Y}^0[f_0]$ is always positive. The threshold of stability 
is thus given by solving the implicit equation
\begin{equation}
\oint |\mbox{d}q|\frac{\sin^2 q}{\sqrt{U_c-\kappa \cos^2 q}} =2\oint |\mbox{d}q|\sqrt{U_c-\kappa \cos^2 q} . 
\label{eq:stabthres}
\end{equation}
The numerical resolution of the above equation for $\kappa=0.3$ yields the value $U_c\approx 0.498$, in 
excellent agreement with the energy value at which Re$(\lambda)$ vanishes (see Fig.~\ref{fig:lyaphmfc2}). 

We note that the above derivation of the threshold energy $U_c$ corroborates with the result derived 
in Ref.~\cite{HMFC2}, where the same result was obtained by developing the single-particle distribution 
as a sum of derivatives of Dirac distributions. The truncation of the expansion to the very first 
term allowed the authors of Ref.~\cite{HMFC2} to obtain the same implicit equation (\ref{eq:stabthres}). 
The approach presented here is more general, since it provides a dispersion relation 
for any stationary distribution, and allows us to derive the stability condition without any additional
hypothesis.

We devote the final part of this Section to the derivation of the growth rate of the instability and
of the threshold energy for the HMF model, in the limit where the on-site 
potential is turned off ($\kappa=0$). Although this result was already obtained \cite{inagaki,choichoi}, its derivation
in this new context allows us to point out the connection between action-angle variables $(\phi,J)$ and 
the canonical ones $(q,p)$. In fact, when only the mean-field potential couples the particles, the 
non-magnetized inhomogenerous stationary states become homogeneous in $q$ and, in correspondence, 
the action-angle variables reduce, modulo a sign, to the canonical coordinates
\begin{eqnarray} 
J&=&\frac{1}{2\pi}\oint p(h,q)\, \mbox{d}q=|p|,
\\ \omega&=&\frac{\displaystyle \partial h}{\displaystyle\partial J}=|p|,
\\ \phi &=&\omega\int^q \frac{\mbox{d}q'}{p}=\mbox{sign}(p)\,q.\label{eq:phijhmf}
 \end{eqnarray}
The presence of absolute values is due to the fact that the action-angle variables take into account 
the direction of the motion along the trajectories, which are now ballistic.
Then, inserting the following relations
\begin{eqnarray}
\int^q e^{\lambda {q'}/{p}}\sin q'\mbox{d}q' &=& \frac{e^{\lambda{q}/{p}}}{1+{\lambda^2}/{p^2}}\left(\frac{\lambda}{p}\sin q-\cos q\right),
\\ \int^q e^{\lambda{q'}/{p}}\cos q'\mbox{d}q' &=& \frac{e^{\lambda{q}/{p}}}{1+{\lambda^2}/{p^2}}\left(\sin q+\frac{\lambda}{p}\cos q\right),
\end{eqnarray}
into Eq.~(\ref{eq:IXY}), one can explicitely write the dispersion relation (\ref{eq:disprelhmfc2}) as
\begin{equation}
\left(1+\pi \int \mbox{d}p \frac{f'_0(p)}{p\left(1+\frac{\lambda^2}{p^2}\right)} \right)^2
+\left(\lambda\pi \int \mbox{d}p \frac{f'_0(p)}{p^2\left(1+\frac{\lambda^2}{p^2}\right)}\right)^2=0.\label{eq:disphmf3}
\end{equation}
The waterbag distribution is now homogeneous in $q$ and symmetric in $p$
\begin{equation}
f_{0}(p)=\frac{1}{2\pi}\frac{1}{2\Delta p}\left(\Theta(p+\Delta p)-\Theta(p-\Delta p)\right),
\end{equation}
and its derivative is given by
\begin{equation}
f'_{0}(p)=\frac{1}{2\pi}\frac{1}{2\Delta p}\left(\delta(p+\Delta p)-\delta(p-\Delta p)\right).
\end{equation}
The second quadratic term in Eq.~(\ref{eq:disphmf3}) vanishes, and one obtains
\begin{equation}
0=1+\pi \int dp \frac{f'_0(p)}{p\left(1+\frac{\lambda^2}{p^2}\right)}=1-\frac{1}{2\Delta p^2\left(1+\frac{\lambda^2}{\Delta p^2}\right)}.
\end{equation}
We finally obtain the complex growth rate
\begin{equation}
\label{eq:growthHMF}
\lambda=\pm\sqrt{\frac{1}{2}-\Delta p^2}~,
\end{equation}
which shows that the waterbag is stable beyond the threshold energy $U_c=1/12$, since
the energy of the system is given by $U=\frac{\Delta p^2}{6}$.

\begin{figure}[!ht]
\centerline{
$\begin{array}{c}
\epsfig{figure=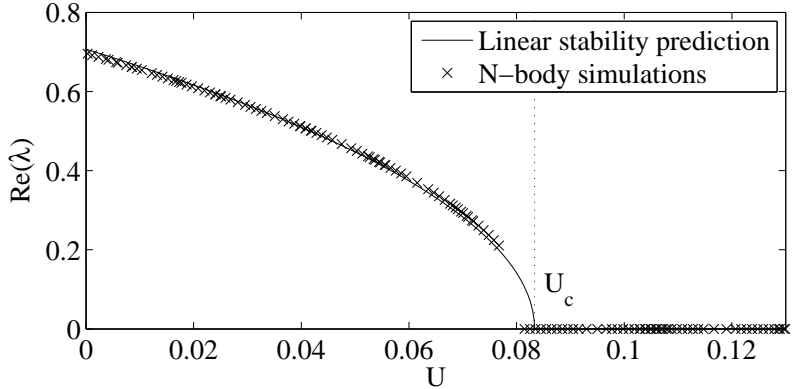,width=12cm}
\end{array}$}
\caption{Growth rate Re$(\lambda)$ of the instability (full line) as a function of the energy $U$ for the HMF
model (model (\ref{eq:HMFcos2}) with $\kappa=0$), as obtained analytically in formula (\ref{eq:growthHMF}). The crosses are
the results of exponential fits of the short-time evolution of the magnetization for the 
$N$-body HMF Hamiltonian.\label{fig:lyaphmf}}
\end{figure}

Fig.~\ref{fig:lyaphmf} shows the comparison of this analytical prediction with the numerical results obtained
for the $N$-body simulations of the HMF model: the agreement is excellent.

\section{The mean-field $\varphi^4$ model\label{sec:phi4}}

The second example that we consider is the mean-field $\varphi^4$ model introduced by 
Desai and Zwanzig~\cite{Zwanzig}. It is a system where the particles are trapped in an external 
double-well potential, and are in addition coupled via a infinite-range force.
It is described by the following Hamiltonian
\begin{equation}
H[f]=\iint \mbox{d}q\mbox{d}p\, f(q,p) \left[\frac{p^2}{2} + \left(\frac{q^4}{4}-(1-\theta)
\frac{q^2}{2}\right) - \frac{\theta}{2} q \iint \mbox{d}q' \mbox{d}p' f(q',p')\ q' \right].
\end{equation}
Notice that positive (resp. negative) values of the parameter $\theta$ correspond to attractive 
(resp. repulsive) mean-field forces. We have used the same parametrization introduced in Ref.~\cite{Zwanzig},  
which can be shown to be minimal by conveniently rescaling the variables and time.  The magnetization $M$ 
is now defined as $M[f]=\iint \mbox{d}q\mbox{d}p\, f(q,p)\,q$, so that the mean-field potential is 
given by $V[f](q)=-(\theta/2) q M[f]$, whereas the external potential is $W(q)=q^4/4-(1-\theta)q^2/2$. It
displays a double well for $\theta < 1$ and a single well otherwise. The solution in the canonical ensemble 
has been recently derived in Ref.~\cite{phi4}, emphasizing that the system exhibits a second order phase transition.
When $\theta=1/2$, the critical temperature has been found to be $T_c\simeq0.264$, corresponding to a critical 
energy $U_c^*=T_c/2\simeq0.132$. The model has been also solved in the microcanonical ensemble and
the entropy as a function of energy and magnetization has been derived using large deviations~\cite{Campa,Hahn,Touchette},
giving equivalent results. However, it has been shown that, in the microcanonical ensemble, magnetic
susceptibility can be negative~\cite{Touchette, Campa}.

\begin{figure}[!ht]
\centerline{
$\begin{array}{cc}
\epsfig{figure=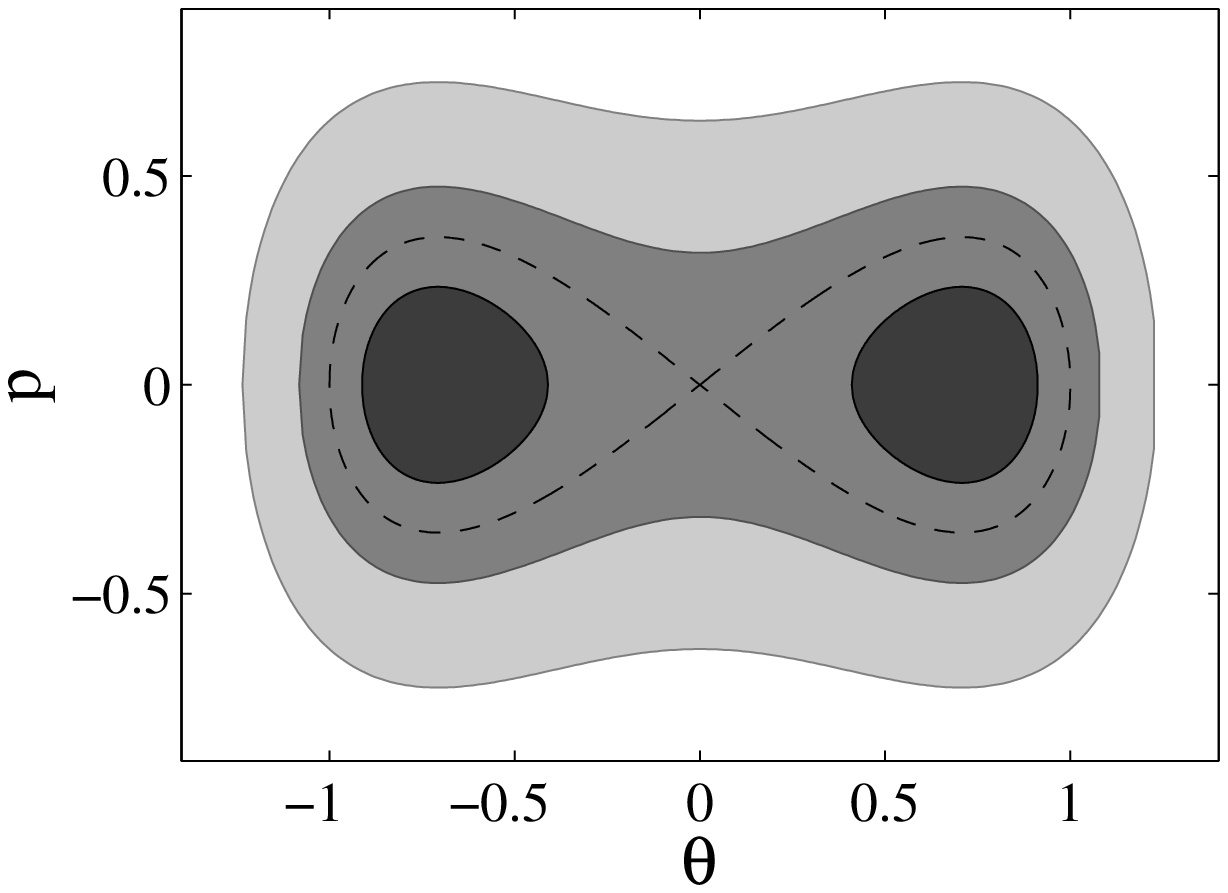,width=8cm}&\epsfig{figure=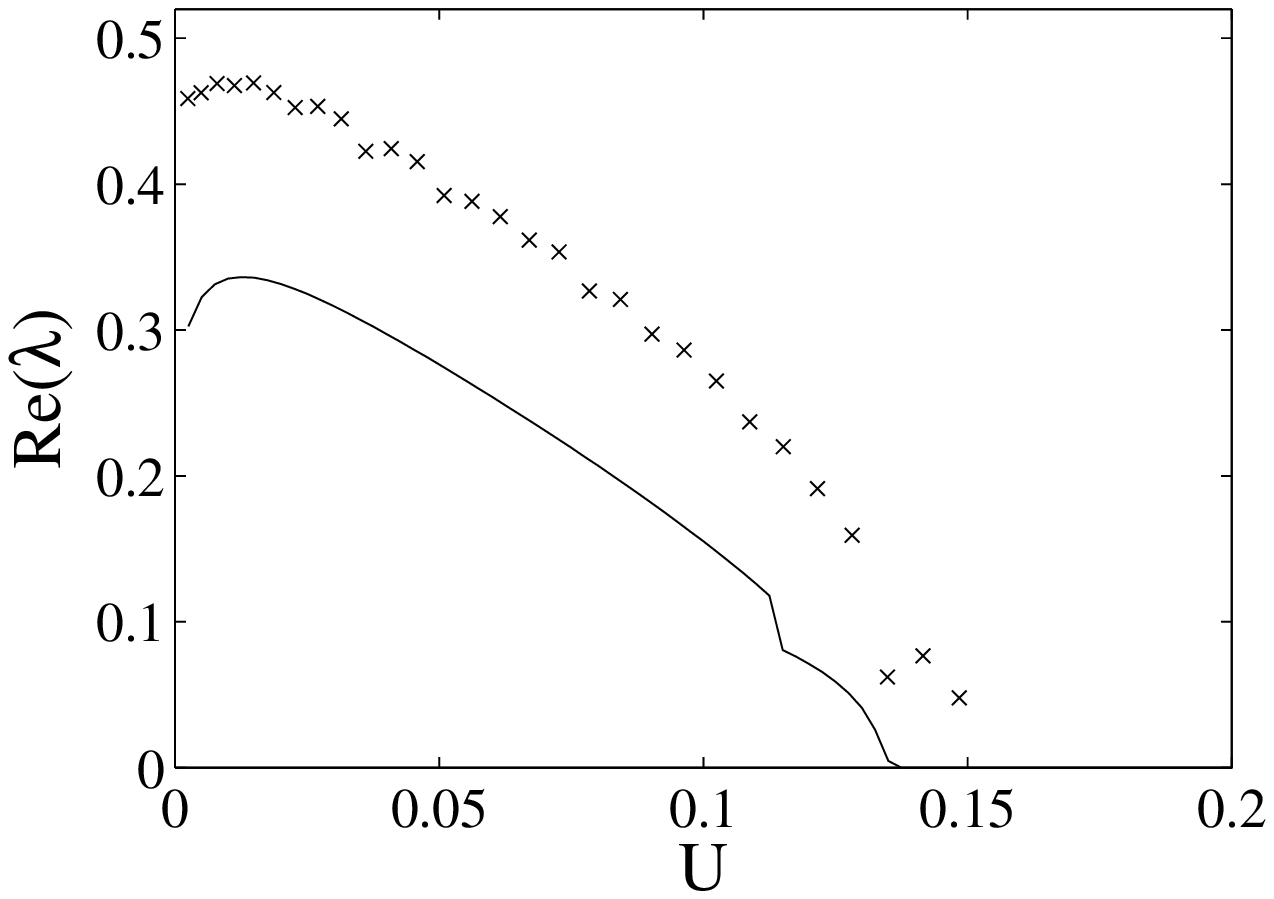,width=8cm}
\end{array}$}
\caption{Left panel: Representation in the $(q,p)$ plane of the waterbags in action-angle. The boundary energies
are $U=-0.035$, $0.05$ and $0.2$ for lighter and lighter grey levels. The dashed line corresponds to 
the separatrix, which has energy $h=0$. Right panel: Growth rate $Re(\lambda)$ (full line) computed by numerically solving 
Eq.~(\ref{eq:dispphi4-2}). The crosses represent the short-time exponential rate of growth of the magnetization
obtained in numerical simulations of the $N$-body Hamiltonian with $N=10^6$.\label{fig:phi4}}
\end{figure}

For this system, the dispersion relation (\ref{eq:fbar2}) takes the following form
\begin{equation}
\bar{f}+\theta M[\bar{f}]f'_0(J)\frac{e^{-\lambda{\phi}/{\omega(J)}}}{\omega(J)} q\int_{0}^{q}
e^{\lambda{\phi'}/{\omega(J)}} \mbox{d}q'=0.
\label{eq:dispphi4}
\end{equation}
The magnetization $M[\bar{f}]$ can be factored out by multiplying this latter expression by $q$ and 
by integrating it over the phase-space. One gets
\begin{equation}
1+\theta \int \mbox{d}J f'_0(J)\oint \mbox{d}\phi \frac{e^{-\lambda{\phi}/{\omega(J)}}}{\omega(J)} 
q^2\int_{0}^{q}e^{\lambda{\phi'}/{\omega(J)}} \mbox{d}q'=0.\label{eq:dispphi4-2}
\end{equation}
Before proceeding to the numerical solution of the above dispersion relation, let us derive explicitly 
the expression that allows us to obtain the stability threshold by setting $\lambda=0$ in the previous
formula. The last integral in Eq.~(\ref{eq:dispphi4-2}) gives trivially $q$, while $\mbox{d}\phi/\omega$ 
can be rewritten as $\mbox{d}q/p$ thanks to Eq.~(\ref{eq:dphi}). One finally gets
\begin{equation}
1+\theta \int \mbox{d}J f'_0(J)\oint \frac{q^2}{p} \mbox{d}q=0.\label{eq:stabphi4-1}
\end{equation}
Let us now restrict to those stationary distributions for which the mean-field vanishes, i.e. $M[f_0]=0$.
This case includes those distributions that are symmetric with respect to $q=0$.
For clarity purposes, we shall also restrict to waterbag distributions that have a boundary energy $U>0$,
i.e. $f_0 (J)$ is constant for all actions $0 < J < J(U)$ and zero for $J>J(U)$. Waterbags with both 
positive and negative boundary energy $U$ are shown in Fig.~\ref{fig:phi4}.

By introducing the following set of variables
\begin{eqnarray}
q &=& x\bar{q},
\\ \bar{q}&=&\sqrt{\sqrt{4h+(1-\theta)^2}-(1-\theta)},
\\ \rho &=&\sqrt{\frac{\sqrt{4h+(1-\theta)^2}+(1-\theta)}{\sqrt{4h+(1-\theta)^2}-(1-\theta)}}~,
\end{eqnarray}
the momentum of a particle with positive energy $h$ can be written as
\begin{equation}
p=\pm \sqrt{2(h-W(q))}
=\pm \frac{\bar{q}^2}{\sqrt{2}}\sqrt{(\rho^2-x^2)(1+x^2)}~.
\end{equation}
Note that $x$ varies in the range $[-\rho;\rho]$, so that the maximum position along a trajectory 
is~$\rho \bar{q}$. Now, the action-angles variables (\ref{eq:j},\ref{eq:phi}) assume the following form
\begin{eqnarray}
J&=&\frac{\bar{q}^3}{2\sqrt{2}\pi}\oint \sqrt{(\rho^2-x^2)(1+x^2)}\, \mbox{d}x 
\\ &=&\frac{\bar{q}^3\sqrt{2}}{3\pi}\left[(\rho^2-1)\mathscr{E}(-\rho^2)+(\rho^2+1)\mathscr{K}(-\rho^2)\right],\label{eq:jphi4}
\\ \omega^{-1}&=& \frac{\sqrt{2}}{2\pi\bar{q}}\oint \frac{\mbox{d}x}{\sqrt{(\rho^2-x^2)(1+x^2)}}=\frac{2\sqrt{2}}{\pi\bar{q}}\mathscr{K}(-\rho^2),\label{eq:omegaphi4}
\\ \phi &=& \omega \int_{0}^q \frac{\mbox{d}x}{\sqrt{(\rho^2-x^2)(1+x^2)}}=\omega \mathscr{F}\left(\frac{x}{\rho},-\rho^2\right).\label{eq:phiphi4}
\end{eqnarray}
 
Using the following relation
\begin{eqnarray}
\oint \frac{q^2}{\sqrt{2(h-W(q))}} |\mbox{d}q| &=&\sqrt{2}\bar{q}\oint \frac{x^2 \mbox{d}x}
{\sqrt{(\rho^2-x^2)(1+x^2)}}
\\ &=& 4\sqrt{2}\bar{q}\left[\mathscr{E}(-\rho^2)-\mathscr{K}(-\rho^2)\right],
\end{eqnarray}
one can show, taking also Eqs.~(\ref{eq:jphi4}) and (\ref{eq:omegaphi4}) into account, that
\begin{equation}
\oint \frac{q^2}{\sqrt{2(h-W(q))}} |\mbox{d}q|=\frac{12 \pi J}{\bar{q}^2(\rho^2-1)}
-\frac{4\pi\bar{q}^2\rho^2}{(\rho^2-1)\omega}.
\end{equation}
Considering that $\bar{q}^2(\rho^2-1)=2(1-\theta)$ and $\bar{q}^2\rho^2/(\rho^2-1)=2h/(1-\theta)$, 
we eventually get the following expression for the stability threshold
\begin{equation}
1+\frac{2\pi \theta}{1-\theta} \int_{J_0}^\infty f_0'(J)  \left(3J-4\frac{h(J)}
{\omega(J)}\right) \mbox{d}J=0.\label{eq:stabphi4-2}
\end{equation}
Let us now consider the case of the waterbag defined by Eq.~(\ref{eq:wb}) with $J_1=0$ and $J_2=J(U)$.
The dispersion relation for this waterbag reads
\begin{equation}
1-\frac{\theta}{1-\theta}\left(3-4\frac{U}{\omega(U)J(U)}\right)=0.\label{eq:stabphi4-3}
\end{equation}
Solving numerically this latter equation for $\theta=1/2$ gives the threshold energy $U_c \simeq 0.144$,
which turns out to be pretty close to the value of the statistical transition energy $U_c^*$ found 
in Ref.~\cite{phi4}.
We can also solve numerically the dispersion relation (\ref{eq:dispphi4-2}) and obtain
the growth rate Re$(\lambda)$. In Fig.~\ref{fig:phi4} this growth rate is compared to a fit
of the short-time exponential growth of the magnetization obtained by integrating numerically
the $N$-body Hamiltonian. Unfortunately the agreement is only qualitative, although the stability threshold 
is correctly reproduced.

\section{Concluding remarks\label{sec:ccl}}

Systems with mean-field interactions are well described by the Vlasov equation in the $N \to \infty$ limit.
An infinity of stationary states exists for such equation and the study of their stability is a subject of
paramount importance. Many exact results about {\it homogeneous} stationary appeared in the literature and several stability criteria have been applied. Also, inhomogeneous states have been treated, but the study of their stability is more complex~\cite{campachacha,chachacha,HMFC2,chacha07,chacha10,barre10}. Characterizing analytically the stability of stationary
solutions of the Vlasov equation will have an impact also on the characterization of the slow convergence
to equilibrium observed in systems with long-range interactions \cite{Campa,Leshouches1,Assisi,Leshouches2,bouchet,chavareview},
in particular on the study of Quasi-Stationary-States (QSS), which are ubiquitous long-lived states in the $N$-body dynamics
of long-range systems. It has been shown that the lifetime of QSS diverges algebraically with $N$ in some 
simple models and it has been conjectured that this can happen only when the QSS corresponds to a stable stationary
state of the Vlasov equation \cite{Campa} (see also Ref.~\cite{Caglioti} for an interesting mathematical result along
this direction). Again, most of the studies on QSS are for the {\it homogeneous} case.

In this paper, we have discussed a class of models where, besides the mean-field interaction, particles are
subjected to an external potential. The effect of the external potential is that of creating an inhomogeneity
in the spatial distribution. Hence, these models are naturally endowed with inhomogeneous stationary
states. After rewriting the Vlasov equation in action-angle variables, we have shown that some of this
inhomogeneous stationary states in conjugate coordinates transform into homogeneous stationary
states that are homogeneous in angle.
We have therefore applied the standard tools of linear stability of the Vlasov
equation to derive a dispersion relation, given in formula (\ref{eq:fbar2}), which is the key result of this
paper. We have specialized this formula for two models: the HMF model with additional asymmetry \cite{HMFC2} 
and the mean-field $\varphi^4$ model
\cite{phi4,Hahn,Touchette,Zwanzig}. For these two models it is possible to further simplify the
dispersion relation and to obtain implicit equations that, solved numerically, give both the growth rate of
the instability and the stability threshold. When the real part of the growth rate vanishes, the state is
a stable stationary state of the Vlasov equation. We have checked these results against the numerical
simulation of the Hamiltonian dynamics of the corresponding $N$-body system. The stability thresholds are in general
in good agreement with the theoretical predictions, but for the growth rate the agreement is only qualitative for the $\varphi^4$ model. A case in which the growth rate turns out to be in perfect agreement
with the simulations is the one of the HMF model \cite{inagaki,pichon,hmf95}. 

Those inhomogeneous stationary states that are also stable are good candidates to become QSS at finite $N$.
We have therefore pointed out the existence of a new class of inhomogeneous QSS, for which it will be possible 
in the future to study the law of divergence of the lifetime with system size.

\ack
SR thanks UJF-Grenoble and ENS-Lyon for financial support and hospitality. He also acknowledges 
the financial support of the COFIN07-PRIN program "Statistical physics of strongly correlated systems at and out 
of equilibrium" of the Italian MIUR and of INFN, and of the ANR-10-CEXC-010-01,
{\it Chaire d'Excellence}. This work was carried out in part while S.R. was Weston Visiting
Professor at the Weizmann Institute of Science.


\section*{References}

\begin{thebibliography}{99}

\bibitem{Campa} Campa A, Dauxois T, Ruffo S, 2009 {\it Phys. Rep.} {\bf 480}, 57 

\bibitem{Leshouches1}  Dauxois T, Ruffo S, Arimondo E, Wilkens M (Eds.), 2002 {\it Dynamics and Thermodynamics of Systems with Long-Range Interactions}, {\it Lecture Notes in Physics} {\bf 602}, Springer  

\bibitem{Assisi} Campa A, Giansanti A, Morigi G and Sylos Labini F (Eds.), 2008 {\it Dynamics and Thermodynamics of systems with long range interactions: theory and experiments}, ({\it AIP Conference proceedings} {\bf 970})

\bibitem{Leshouches2} Dauxois T, Ruffo S, Cugliandolo L (Eds.), 2009 {\it Long-Range Interacting Systems}, {\it Lecture Notes of the Les Houches Summer School: Volume 90, August 2008}, Oxford University Press

\bibitem{bouchet} Bouchet F, Gupta S and Mukamel D, 2010 {\it Physica A} {\bf 389} 4389

\bibitem{chavareview} P. H. Chavanis, 2006 {\it Int. J. Mod. Phys. B} {\bf 20} 3113

\bibitem{eebook} Elskens Y and Escande D F, 2003 {\it Microscopic Dynamics of Plasmas and Chaos}, IoP Publishing, Bristol

\bibitem{colsonbonifacio} Colson W, 1976 {\it Phys. Lett. A} {\bf 59} 187; Bonifacio R, Casagrande F and Pellegrini C, 1987 {\it Opt.
Commun.} {\bf 61} 55.

\bibitem{CARL} Bonifacio R and De Salvo Souza L, 1994 {\it Nucl. Instrum. and Meth.} {\bf 341} 360; Bonifacio R, 
De Salvo Souza L, Narducci L, and D'Angelo E J, 1994 {\it Phys. Rev. A} {\bf 50} 1716

\bibitem{TWT} Dimonte G and Malmberg J H, 1978 {\it Phys. Fluids} {\bf 21} 1188; Tsunoda SI, Doveil F, Malmberg JH, 
1987 {\it Phys. Rev. Lett.} {\bf 58}.

\bibitem{inagaki} Inagaki S and Konishi T, 1993 {\it Publ. Astron. Soc. Jpn.}  \textbf{45}, 733. 

\bibitem{pichon} Pichon C, 1994 PhD Thesis Cambridge

\bibitem{hmf95} Antoni M and Ruffo S, 1995 {\it Phys. Rev. E} {\bf 52}, 2361

\bibitem{MesserSpohn} Messer J and Spohn H, 1982 {\it J. Stat. Phys.} {\bf 29}, 561

\bibitem{Braun} Braun W and Hepp K, 1977 {\it Commun. Math. Phys.} {\bf 56} 101

\bibitem{Landaukinetic} Landau L D, 1946 {\it J. of Phys. (USSR)} {\bf 10}, 25

\bibitem{Nicholson} Nicholson D R, 1983 {\it Introduction to Plasma Theory}, John Wiley

\bibitem{Balescubook} Balescu R., 1997 {\it Statistical dynamics: Matter out of equilibrium}, Imperial College
Press, London.

\bibitem{yamaguchi04} Yamaguchi Y Y, Barr\'e J, Bouchet F, Dauxois T and Ruffo S, 2004 {\it Physica A} {\bf 337} 36

\bibitem{BinneyTremaine} Binney J, Tremaine S, 1987 {\it Galactic Dynamics}, Princeton Series in Astrophysics

\bibitem{brink} Brink D M  and Dellafiore A, 1986 {\it Nuclear Physics} {\bf A456} 205

\bibitem{bertin} Bertin G, Pegoraro F, Rubini F and Vesperini E, 1994  {\it Astrophys. J.}  {\bf 434} 94

\bibitem{schindler} Schindler K, 2007 {\it Physics of Space Plasma Activity}, Cambridge University Press

\bibitem{krall}  Krall N A and Trivelpiece A W, 1973 {\it Principles of plasma physics}, McGraw-Hill, New York 

\bibitem{kaljnas} Kalnajs A J, 1971 {\it ApJ} {\bf 166} 275

\bibitem{weinberg} Weinberg  M D, 1989 {\it MNRAS} 239

\bibitem{camporeale} Camporeale E, Delzanno G L, Lapenta G and Daughton W, 2006 {\it Phys. Plasmas}{\bf 13}, 092110

\bibitem{guo} Guo Y and Strauss W A, 1995 {\it Commun. Pure Appl. Math.} {\bf 63} 861

\bibitem{manfredi} Manfredi G, Bertrand P, 2000 {\it Phys. Plasmas} {\bf 7}, 2425

\bibitem{lin2001} Lin Z, 2001 {\it Math. Research Lett.} {\bf 8} 1

\bibitem{lin2005} Lin Z,  2005 {\it Comm. Pure. Appl. Math.} {\bf 58} (4) 505

\bibitem{campachacha} Campa A and Chavanis P-H, 2010 {\it J. Stat. Mech.: Theory and Experiment} P06001

\bibitem{chachacha} Chavanis P. H., arXiv:1007.4916

\bibitem{phi4} Dauxois T, Lepri S and Ruffo S, 2004 {\it Comm. in Nonlin. Sc. and Num. Simu.} {\bf 8}, 375

\bibitem{Hahn}Hahn I and Kastner M, 2006 {\it European Physical Journal B} {\bf 50} 311

\bibitem{Touchette} Campa A, Ruffo, S and Touchette H, 2007 {\it Physica A} {\bf 369}, 517

\bibitem{HMFC2} Jain K, Bouchet F and Mukamel D, 2007 {\it J. Stat. Mech.: Theory and Experiment} {\bf 11}, 8

\bibitem{Zwanzig} Desai R C and Zwanzig R, 1978  {\it J. Stat. Phys.} {\bf 19}, 1

\bibitem{chacha07} Chavanis P-H, 2007 {\it Physica A} {\bf 377}, 469

\bibitem{chacha10} Chavanis P-H, 2010 {\it J. Stat. Mech.: Theory and Experiment} P05019

\bibitem{barre10} Barr\'e J, Olivetti A and Yamaguchi Y Y, 2010 {\it J. Stat. Mech.: Theory and Experiment} P08002

\bibitem{kac} Kac M, Uhlenbeck G E, and Hemmer P C, 1963 {\it J. of Math. Phys.} {\bf 4}, 216

\bibitem{gibbons} Gibbons J, Holm DD, and Tronci C, 2008 {\it Phys. Lett. A} {\bf 372}, 1024

\bibitem{choichoi} Choi M Y and Choi J, 2003 {\it Phys. Rev. Lett.} \textbf{91}, 124101

\bibitem{lichtenberg} Lichtenberg A J and Lieberman M A, 1992 {\it Regular  and Chaotic Dynamics} 2nd Ed., Springer-Verlag, Appli. Math. Sci. {\bf 38}, New York

\bibitem{atela} Mclachlan R I and Atela P, 1992 {\it Nonlinearity} {\bf 5}, 541

\bibitem{Caglioti} Caglioti E, Rousset F., 2007 {\it J. Stat. Phys.} {\bf 129}, 241

\end{thebibliography}
\end{document}